\documentstyle[preprint,pra,aps]{revtex}

\begin{document}

\title{Polarization induced by charged particles in real solids}
\author{I. Campillo$^{1}$ and J. M. Pitarke$^{1,2}$}
\address{$^1$ Materia Kondentsatuaren Fisika Saila, Zientzi Fakultatea, 
Euskal Herriko Unibertsitatea,\\ 644 Posta kutxatila, 48080 Bilbo, Basque 
Country,
Spain\\
$^2$Donostia International Physics Center (DIPC) and Centro Mixto
CSIC-UPV/EHU,\\
Donostia, Basque Country, Spain}

\date\today

\maketitle

\begin{abstract}
We report a first-principles description of the induced wake potential and
density in real solids. The linear-response formalism is used to
obtain the potential and density induced by an external charge penetrating
through an inhomogeneous periodic system. The linear dynamical response of 
the system is evaluated in the random-phase approximation, by including the full
band structure of the solid. The impact of the periodic crystalline potential is
analyzed, and variations of the wake along different channels in Al and Si are
investigated. 
\end{abstract}

\pacs{PACS numbers: 71.45.Gm, 78.47.+p}

\section{Introduction}

A charged particle penetrating a solid causes a distortion 
in the electronic density around the particle and behind its position:
this is what Bohr called the induced density wake\cite{Bohr}. Related to
this induced density there is an induced potential. For a sufficiently high 
velocity the induced wake shows an oscillatory behaviour. When the velocity 
of the projectile is larger than the average velocity of target electrons
(typically $v_F$, the Fermi velocity), one may consider a linear response of
the medium. However, in the case of projectiles moving with smaller velocities,
nonlinearities may play an important role for metallic densities 
($2<r_s<6$, $r_s$ being the average electron distance\cite{note1}).

Much research has been oriented to the study of these quantities. Pioneering
work on dynamic screening was performed by Neufeld and Ritchie\cite{Neufeld}.
They evaluated the induced potential and density wake by using a local 
dielectric function as the linear response function of the medium. Based upon
a linear response of the target, different aspects related to the induced
polarization such as wake riding states\cite{neelavathi,day}
 or the spatial distribution
of the induced potential and density\cite{EtxeRB,mazarro}
 have been investigated. The 
induced potential and density have been studied beyond linear-response
theory, in the static electron gas approximation\cite{esbensen} and within
hydrodynamical formulations\cite{arnau,dorado,Ber1}. The 
quadratic induced polarization
by an external charge in the full random-phase approximation (RPA)
 has been recently reported\cite{Pitt1,bergara1}.

In previous works the induced potential and density have been evaluated
on the basis of a jellium model of the target, which consists of 
an isotropic homogeneous electron gas embedded in a uniformly distributed
positive background. However, in a more realistic approach valence electrons
move in a periodic potential and one-electron excitations split into
the so-called energy bands. The impact of band-structure effects on
plasmon dispersion curves\cite{Quong,Arya,Eguiluzcesio}, 
dynamical structure factors\cite{Godby1,Godby2,Fleszar,Rubio}, 
stopping power\cite{Igorstopprb}, and hot-electron 
lifetimes\cite{Igorprlprb,Ekardt} has
been investigated only very recently.
It has been shown that these effects can be important even in the case 
of free-electron metals such as aluminum. The wake 
potential induced by swift protons through  different solids
has been studied using a model dielectric function\cite{abril}. 

In this paper we report, within linear-response theory, 
an {\em ab initio} evaluation of the induced potential 
and density of ions moving through real solids. Within our description
we can obtain the induced potential under channeling conditions, i. e., 
when the projectile penetrates through different symmetry directions of the
solid. In section II we derive explicit expressions for the 
position-dependent induced potential and density. In section III numerical
calculations of the induced potential in Al and Si are presented,
for both random and crystal-symmetry incident directions. In section IV 
the most relevant conclusions are summarized. Atomic units are used 
throughout, i. e., $m_e=e=\hbar=1$.

\section{Theory}

We consider a point particle of charge $Z_1$ moving in an inhomogeneous system
with velocity $\bf v$ and impact vector $\bf b$, such that
\begin{equation}\label{eq1}
\rho^{ext}({\bf r},t)=Z_1\delta({\bf r}-{\bf b}-{\bf v}t).
\end{equation}

For a periodic crystal we can write:
\begin{equation}\label{eq2}
\rho^{ext}_{{\bf G}}({\bf q},\omega)=2\pi Z_1 e^{-{\rm i}({\bf q}+{\bf G})\cdot {\bf b}}
\delta\left[\omega-({\bf q}+{\bf G})\cdot {\bf v}\right],
\end{equation}
where ${\bf G}$ is a reciprocal-lattice vector and ${\bf q}$ lies in the
Brillouin zone (BZ). The external charge induces a density, so that the total 
density variation of the medium is given by the sum 
\begin{equation}\label{eq3}
\rho_{{\bf G}}({\bf q},\omega)=\rho^{ext}_{{\bf G}}({\bf q},\omega)+
\rho^{ind}_{{\bf G}}({\bf q},\omega).
\end{equation}
Poisson's equation allows us to write the potential related to a given density.
In our case,
\begin{equation}\label{eq4}
\phi_{{\bf G}}({\bf q},\omega)=v_{{\bf G}}({\bf q})\rho_{{\bf G}}({\bf q},\omega),
\end{equation}
where $v_{{\bf G}}({\bf q})=4\pi/|{\bf q}+{\bf G}|^2$ and $\phi_{{\bf G}}({\bf q},\omega)$ 
represent the Fourier components of the bare Coulomb potential and the
total potential, respectively. Within linear response theory, we can write the
total potential in terms of the external charge:
\begin{equation}\label{eq5}
\phi_{{\bf G}}({\bf q},\omega)=\sum_{{\bf G}'}\epsilon^{-1}_{{\bf G},{\bf G}'}({\bf q},\omega)
v_{{\bf G}'}({\bf q})\rho^{ext}_{{\bf G}'}({\bf q},\omega).
\end{equation}
$\epsilon^{-1}_{{\bf G},{\bf G}'}({\bf q},\omega)$ are the Fourier coefficients
of the dielectric function, which within the RPA
read
\begin{equation}\label{eq6}
\epsilon_{{\bf G},{\bf G}'}({\bf q},\omega)=\delta_{{\bf G},{\bf G}'}-v_{{\bf
G}'}({\bf q})
\chi^0_{{\bf G},{\bf G}'}({\bf q},\omega),
\end{equation}
where $\chi^0_{{\bf G},{\bf G}'}({\bf q},\omega)$ represent the Fourier coefficients of
the density-response function of non-interacting electrons,
\begin{eqnarray}\label{eq7}
\chi_{{\bf G},{\bf G}'}^0({\bf q},\omega)=&&{1\over \Omega}\sum_{BZ}
\sum_{n}\sum_{n'}
{f_{{\bf k},n}-f_{{\bf k}+{\bf q},n'}\over\varepsilon_{{\bf
k},n}-\varepsilon_{{\bf k}+{\bf q},n'} +(\omega+{\rm i}\eta)}\cr\cr
&&\times\langle\phi_{{\bf k},n}|e^{-{\rm i}({\bf q}+{\bf G})\cdot{\bf
r}}|\phi_{{\bf k}+{\bf q},n'}\rangle
\langle\phi_{{\bf k}+{\bf q},n'}|e^{{\rm i}({\bf q}+{\bf
G}')\cdot{\bf
r}}|\phi_{{\bf k},n}\rangle.
\end{eqnarray}
Here, the sums run over the band structure for each wave vector
${\bf k}$ in the first BZ, $|\phi_{{\bf k},n}\rangle$ and 
$\varepsilon_{{\bf k},n}$ are the one-electron wave functions and energies,
and $f_{{\bf k},n}$ are Fermi factors,
$f_{{\bf k},n}=\theta(E_F-\varepsilon_{{\bf k},n})$. $\Omega$ is the normalization
volume.

Combining Eqs. (\ref{eq3}), (\ref{eq4}) and (\ref{eq5}) we obtain the
following equation for $\rho^{ind}$ in terms of $\rho^{ext}$:
\begin{equation}\label{eq8}
\rho_{{\bf G}}^{ind}({\bf q},\omega)=\sum_{{\bf G}'}\left[ |{\bf q}+{\bf G}|^2
\epsilon^{-1}_{{\bf G},{\bf G}'}({\bf q},\omega)|{\bf q}+{\bf
G}'|^{-2}-\delta_{{\bf G},{\bf G}'}\right]
\rho_{{\bf G}'}^{ext}({\bf q},\omega).
\end{equation}
Substituting Eq. (\ref{eq2}) into Eq. (\ref{eq8}) and Fourier
transforming back to real space yields
\begin{eqnarray}\label{eq9}
\rho^{ind}({\bf r},t)=&&{Z_1\over \Omega}\sum_{{\bf q}}^{BZ}\sum_{{\bf G}}
e^{{\rm i}({\bf q}+{\bf G})\cdot ({\bf r}-{\bf b}-{\bf v}t)}\sum_{{\bf K}}
e^{-{\rm i}{\bf K}\cdot{\bf b}}e^{-{\rm i}{\bf K}\cdot{\bf v}t}\cr\cr
&&\times\left[ |{\bf q}+{\bf G}|^2
\epsilon^{-1}_{{\bf G},{\bf G}+{\bf K}}({\bf q},({\bf q}+{\bf G}+{\bf K})\cdot{\bf v})
|{\bf q}+{\bf G}+{\bf K}|^{-2}-\delta_{{\bf G},{\bf G}+{\bf K}}
\right],
\end{eqnarray}
where we have made use of
\begin{equation}\label{eq10}
\int {{\rm d}^3{{\bf q}}\over (2\pi)^3}\equiv {1\over\Omega}\sum_{{\bf q}}^{BZ}
\sum_{{\bf G}},
\end{equation}
and we have set ${\bf G}'={\bf G}+{\bf K}$.

If we consider a definite trajectory of the projectile, only the ${\bf K}$ vectors 
such that ${\bf K}\cdot{\bf v}=0$ contribute to the sum, and
\begin{eqnarray}\label{eq11}
\rho^{ind}({\bf r},t)=&&{Z_1\over \Omega}\sum_{{\bf q}}^{BZ}\sum_{{\bf G}}
e^{{\rm i}({\bf q}+{\bf G})\cdot ({\bf r}-{\bf b}-{\bf v}t)}|{\bf q}+{\bf G}|^2\sum_{{\bf K}}{'}
e^{-{\rm i}{\bf K}\cdot{\bf b}}\cr\cr
&&\times\left[
{\epsilon^{-1}_{{\bf G},{\bf G}+{\bf K}}({\bf q},({\bf q}+{\bf G}+{\bf K})\cdot{\bf v})\over
|{\bf q}+{\bf G}+{\bf K}|^{2}}-{\delta_{0,{\bf K}}\over |{\bf q}+{\bf G}|^{2}}
\right].
\end{eqnarray}
The prime in the sum over ${\bf K}$ accounts for the ${\bf K}\cdot{\bf v}=0$ condition.
Therefore, Eq. (\ref{eq11}) gives the density induced by an external charged
particle moving in a definite trajectory through a crystal.

In order to get the induced potential we just apply Poisson equation,
which yields
\begin{eqnarray}\label{eq12}
\phi^{ind}({\bf r},t)=&&{-4\pi Z_1\over \Omega}\sum_{{\bf q}}^{BZ}\sum_{{\bf G}}
e^{{\rm i}({\bf q}+{\bf G})\cdot ({\bf r}-{\bf b}-{\bf v}t)}\sum_{{\bf K}}{'}
e^{-{\rm i}{\bf K}\cdot{\bf b}}\cr\cr
&&\times\left[
{\epsilon^{-1}_{{\bf G},{\bf G}+{\bf K}}({\bf q},({\bf q}+{\bf G})\cdot{\bf v})\over
|{\bf q}+{\bf G}+{\bf K}|^{2}}-{\delta_{0,{\bf K}}\over |{\bf q}+{\bf G}|^{2}}
\right].
\end{eqnarray}
We can define an average or induced random potential as the mean value over
impact vectors $\bf b$, which results in the ${\bf K}=0$ term of Eq. (\ref{eq12}):
\begin{equation}\label{eq13}
\phi^{ind}_{random}({\bf r},t)={-4\pi Z_1\over \Omega}\sum_{{\bf q}}^{BZ}\sum_{{\bf G}}
{e^{{\rm i}({\bf q}+{\bf G})\cdot ({\bf r}-{\bf v}t)}\over
|{\bf q}+{\bf G}|^{2}}
\left[\epsilon^{-1}_{{\bf G},{\bf G}}({\bf q},({\bf q}+{\bf G})\cdot{\bf v})-1
\right].
\end{equation}

The most important contribution to the position-dependent induced potential
of Eq. (\ref{eq12}) is provided by the ${\bf K}=0$ term. For those directions
for which there are not reciprocal vectors satisfying the ${\bf K}\cdot{\bf v}=0$
condition, we have the average potential of Eq. (\ref{eq13}). For a few 
highly symmetric or {\em channeling} conditions, non-negligible corrections
to the random result are found. The random induced potential exactly coincides
with the well-known jellium result\cite{Etxe} when one replaces the inverse
dielectric matrix entering Eq. (\ref{eq13}) by the inverse dielectric
function of a homogeneous electron gas\cite{Lind}.

The position-dependent stopping power can be obtained directly from Eq. 
(\ref{eq12}), by simply taking into account that
\begin{equation}\label{eq14}
-{{\rm d}E\over {\rm d}x}={Z_1\over v}\bigtriangledown\phi^{ind}{\big {|}}_{
{\bf r}={\bf b}+{\bf v}t}\cdot{\bf v},
\end{equation}
which gives an expression that  
exactly coincides with the position-dependent stopping
power derived from the knowledge of the imaginary part of the projectile
self-energy\cite{Igorstopprb}.

The main ingredient of our calculation of the induced potential
and density is the inverse dielectric matrix, which we have evaluated
in the RPA by inverting Eq. (\ref{eq6}). Hence, at this point we have only
considered the  average electrostatic interaction between the electrons. We have
not found differences in the induced potential by including many-body
short-range correlations in the response of the system. The one-electron
Bloch states entering Eq. (\ref{eq7}) are the self-consistent LDA
eigenfunctions of the Kohn-Sham equation of density-functional theory
(DFT)\cite{Kohn,Sham}. 
We first expand the states in a plane-wave basis and then 
solve for the coefficients of the expansion self-consistently. The electron-ion
interaction is described in terms of a non-local, norm-conserving ionic
pseudopotential\cite{HSC}. The XC potential is computed with use of 
the energy first calculated by Ceperley and Adler\cite{CA} and then parametrized
by Perdew and Zunger\cite{Perdew}. 
We subsequently evaluate the $\chi^0_{{\bf G},{\bf G}'}({\bf q},\omega)$
polarizability and invert Eq. (\ref{eq6}) as we sum over ${\bf q}$, ${\bf
G}$, and
${\bf K}$ to obtain the induced potential and density.

\section{Results}

In this section
we present the results of our calculations of the induced potential when 
a charged particle penetrates through aluminum and silicon. 
The average electron densities of Al and Si are similar: the
corresponding free-electron gas (FEG) is characterized by $r_s=2.07$ for
Al and $r_s=2.01$ for Si. However, aluminum is a metal and silicon
a semiconductor and the induced potential will exhibit a different behaviour,
which we would not obtain on the basis of a FEG calculation. 

Although Al is usually regarded as a jellium-like material, inelastic 
X-ray scattering experiments\cite{Platzman} 
and theoretical analyses of the dynamical
structure factor\cite{Godby1,Fleszar},  
stopping power\cite{Igorstopprb}, and hot-electron 
decay rates\cite{Igorprlprb} have 
revealed that it is necessary to take into account
the full band structure for a proper description of its electronic properties.
In this work, Bloch states have been expanded in a plane-wave basis with an
energy cutoff of 12 Ry, which corresponds to keeping in this expansion $\sim 100$
plane waves. The sums over the BZ for both the polarizability $\chi^0_{{\bf
G},{\bf G}'}({\bf q},\omega)$ and the induced potential have been performed
on $10\times 10\times 10$ Monkhrost and Pack meshes\cite{MP}. 
The sum over reciprocal-lattice vectors
in the potential has been extended to the first 15 ${\bf G}$ vectors, which
corresponds to a cutoff in the momentum transfer of $2.9q_F$ ($q_F$ is the Fermi
momentum). We have included 30 bands in the sums over the band structure for
each ${\bf k}$ vector in Eq. (\ref{eq7}), which allows us to calculate the
induced potential and the stopping power up to velocities of the  order of 2
a.u.\cite{Igorstopprb}. 

Silicon is a covalent crystal which shows strong valence-electron density
variations in certain directions. This allows the formation of channels
in which the density is very low. 
For example, the integrated 
density in the $\langle 110\rangle$ channel
 varies from $r_s=1.49$ at the atomic row 
 to $r_s=3.37$ at the center of the channel,
i.e., an $80\%$. This has an obvious impact on the induced potential and the
stopping power. The covalent character of Si imposes a higher cutoff
for the Bloch-state expansion than in Al, due to the higher degree of
localization of the electronic states in the former material. 
This also results in Si having bands that are flatter than in the case of
Al, so that the number of bands included in the calculation of the
polarizability 
$\chi^0_{{\bf G},{\bf G}'}({\bf q},\omega)$ must be larger in the case of Si if
the same energy transfers as in Al are to be included. We have used a cutoff of
16 Ry ($\sim 300$ plane waves/state), and have included 100 bands in the
calculation of $\chi^0_{{\bf G},{\bf G}'}({\bf q},\omega)$. The sums over the BZ
have been performed on $8\times 8\times 8$ Monkhorst and Pack meshes\cite{MP},
and the sum over reciprocal-lattice vectors has been extended to the first 15
${\bf G}$ vectors, which corresponds to a momentum-transfer cutoff of $2.1q_F$.

We focus on the spatial distribution of the induced potential along the
incoming particle trajectory.
The $z$ coordinate appearing
in the figures is always relative to the particle position. 
The stopping power derived from the slope of the induced potential at
the projectile position coincides with the stopping power reported in
Refs. \onlinecite{Igorstopprb} and \onlinecite{Igorstopprb2}. An ion of $Z_1=1$
is always considered.

In Fig. 1 we have plotted the random potential induced by an ion with 
$v=0.6$ a.u. through Al (a) and Si (b). The solid line represents the
calculation for the real solid, whereas the dashed line represents the
FEG calculation. For this low-velocity regime (below $v_F$), differences
between full band-structure and FEG calculations come from the presence
of interband transitions and, also, the gap in the case of Si. The 
splitting of the bands makes the polarization easier and, for this
reason, the induced potential at the origin is higher
for the real crystal than for the FEG in Al. In the case of
Si, the slope of the induced potential at the origin is smaller than 
in the corresponding
FEG, due to the presence of the gap, which, for low-energy transfers diminishes
the polarization of the electron system. The gap of 
semiconductors like Si brings about a non-linear stopping power
for low velocities\cite{Igorstopprb2} and an induced potential which
presents, at these velocities, a lower slope.

In Fig. 2 we have plotted the random potential induced by an ion with
$v=1.5$ a.u. through Al (a) and with $v=1.6$ a.u. 
through Si (b). Solid and dashed lines represent full band-structure and FEG
calculations, respectively. These velocities are well over the
plasmon-excitation threshold ($v_t\sim 1.3$ a.u. in both cases). An oscillatory
behaviour appears behind the ion due to plasmon
excitation\cite{Neufeld,EtxeRB,mazarro}, the wavelength of the  oscillations
being of $\sim 2\pi v/\omega_p$ ($\omega_p$ is the  plasma frequency). As can be
appreciated in Figs. 2a and 2b, the wavelength of the random potential is the
same in the real crystal and in the corresponding FEG for both materials. This
is a consequence of the fact that plasmon contributions to the stopping power
are the same for both the real solid and the  FEG\cite{Igorstopprb}. However,
the oscillations in real Si are more damped than in real Al, due to the shorter
lifetime of plasmons in Si which stems from the higher density of bands, thus
increasing the decay channels\cite{Igorstopprb2}.

In Fig. 3 we have plotted the position-dependent potential, Eq. 
(\ref{eq12}), induced by an ion with $v=1.5$ a.u. along the $\langle 100\rangle$
direction in Al (a), and with $v=2.0$ a.u. along the $\langle 110 \rangle$
direction in Si (b). The impact vector is $b(0,1,0)$ for Al and $b(-1,1,0)$ for
Si. 
$b$ is measured in units of the lattice constant $a_c$ of the target.
Calculations
for an impact parameter $b=0$ (atomic row) and $b=0.25$ 
(center of the channel) 
are represented by  solid and short-dashed lines, respectively. For comparison,
a local-density approximation (LDA)\cite{note2} of the induced
potential is also displayed in Fig. 3 by a dashed line for $b=0$ 
and a dotted line for $b=0.25$. In
this approach the position-dependent induced potential is obtained as the
potential induced in a FEG with an electron density equal to the average
electron density along the projectile path. 
The slope of the induced potential gives the stopping power in each case. We
can appreciate a substantial variation in the slope of Si as we move
across the channel, due to the great variation of the density. A
lower density at the center of the channel results in a smaller slope and 
a lower stopping power than in the atomic row and in the random case. 
Variations in the spatial distribution of the induced potential 
in Al are fairly small as $b$ changes, because the electron density
is almost flat in this material.  
If we calculated 
the potential at the ion position 
 as a function of $b$, we would obtain a qualitatively similar result in
both the {\em ab initio} and FEG evaluation, i.e., following the local
density variation\cite{Igorstopprb}. 
However, it is clear from Figs 3a and 3b that the
spatial distribution of the induced potential is not well described by 
the LDA calculation. Above all in Si, Fig. 3b, the differences are more
dramatic. The oscillations behind the ion have the same wavelength 
independently of $\bf b$ in the crystal calculation, whereas in the 
LDA calculation the wavelength depends on the average density for each
path. Plasmon contributions appear in the ${\bf K}=0$ term of the 
position-dependent induced potential of Eq. (\ref{eq12}). The remaining terms
(${\bf K}\neq 0$) are mainly corrections to the single particle e-h 
contribution, which are due to the presence of density variations away from the
projectile path, i.e., the so-called crystalline local-field corrections.
For this reason, the wavelength of the induced potential remains unchanged.
Furthermore, the stopping-power peak 
is located at the same position for any velocity direction and for any 
impact
parameter\cite{Igorstopprb2}, since plasmons are collective excitations 
which involve all the electrons of the system. However,
in an LDA calculation the stopping-power peak
is located at the velocity that corresponds to the average density of the 
path and not to the average target density.
These results lead us to the conclusion that LDA calculations are not 
suitable for the calculation, at these velocities, of the position-dependent
induced potential and stopping power.

\section{Conclusions}

We have presented full band-structure calculations of both random and 
position-dependent induced potentials in Al and Si. The linear-response 
formalism has been used to obtain the potential and density induced by an
external charge penetrating through an inhomogeneous periodic system.
The random potential has been evaluated in the RPA for velocities
below and above the plasmon-excitation threshold. For low velocities,
differences between the FEG and {\em ab initio} calculations come 
from the sensitiveness to the band structure of the
target. For higher velocities, we have shown that oscillations 
behind the ion have the same
wavelength in both the FEG and the real crystal, due to the 
fact that plasmon excitation remains unchanged. Finally, 
we have investigated the position-dependent potential induced by projectiles
incident along the $\langle 100\rangle$ direction in Al and along the widest
channel in Si, the
$\langle 110\rangle$ direction. Variations in the spatial distribution of the
induced potential are more pronounced in the case of Si. Besides, we have shown
that the LDA calculation does not properly account, at the velocities under
study, for the spatial distribution of the induced potential along the channel.

\section{Acknowledgments}

We thank P. M. Echenique and A. G. Eguiluz for stimulating discussions.
We acknowledge partial support by the University of the Basque Country,
the Basque Unibertsitate eta Ikerketa Saila, and the Spanish Ministerio de
Educaci\'on y Cultura.

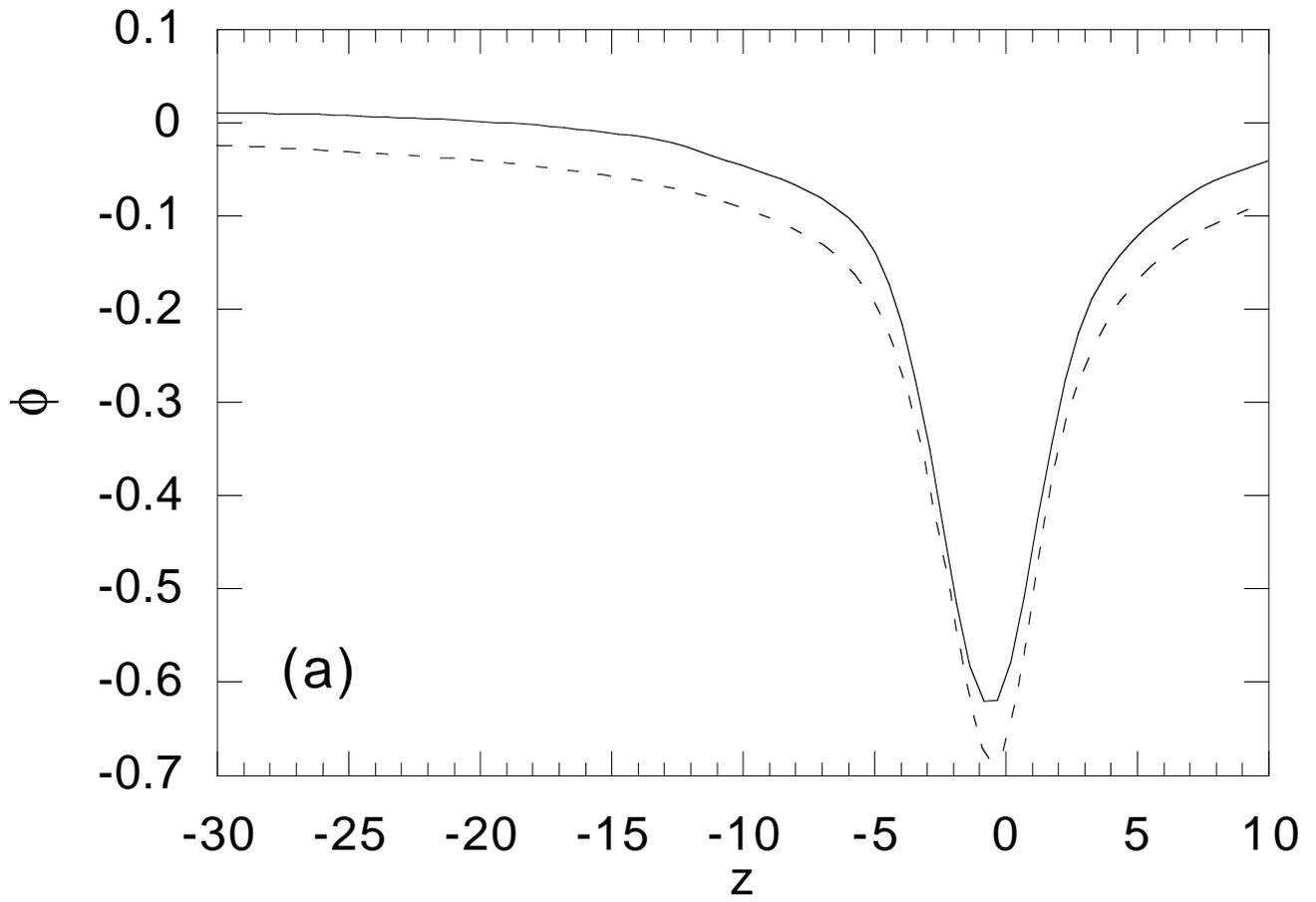
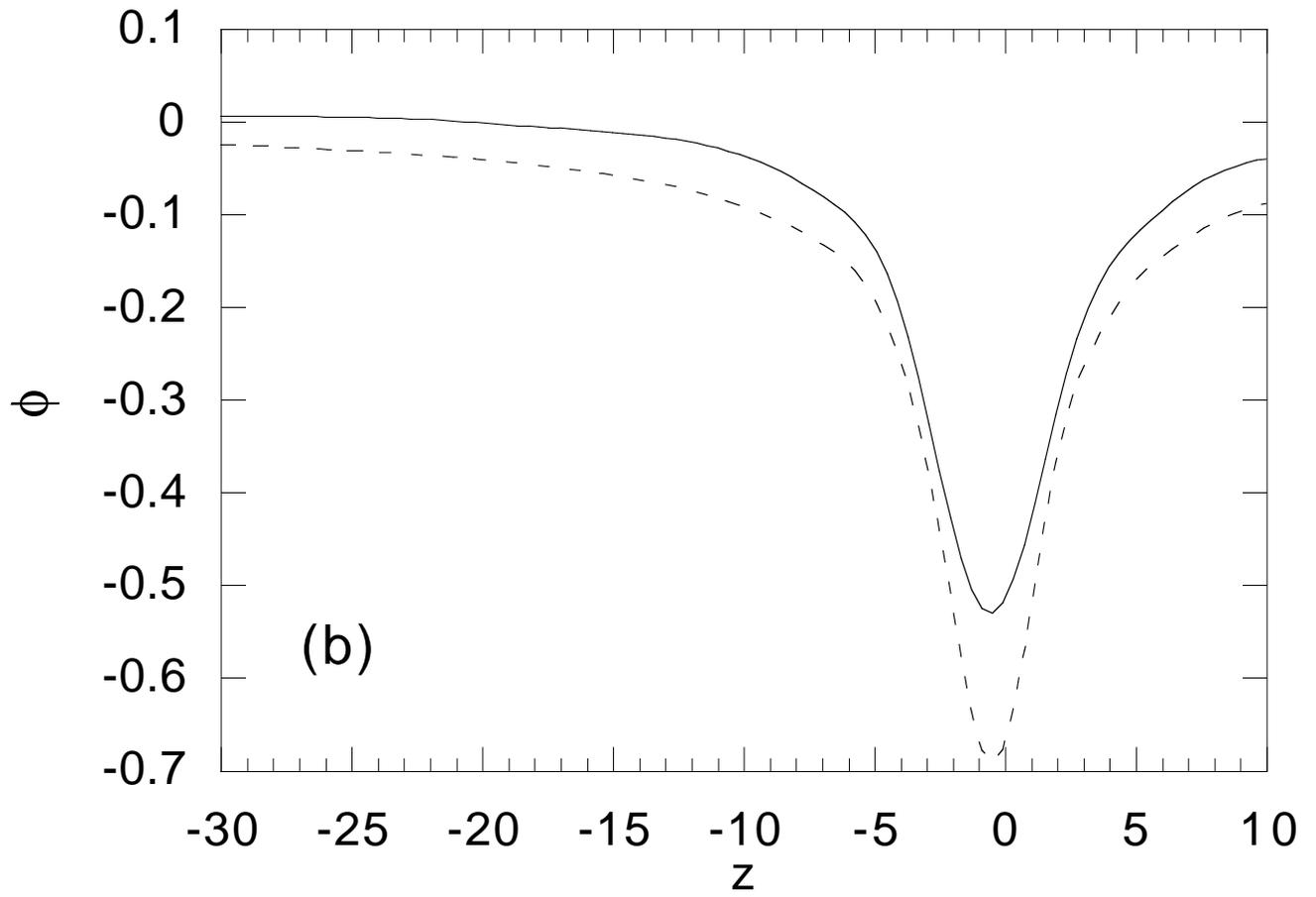
\begin{figure}
\caption[]{Random potential induced by a proton ($Z_1=1$) moving with $v=0.6$
a.u. through Al (a), and Si (b). Solid and dashed lines represent full
band-structure and FEG calculations, respectively.}
\end{figure}

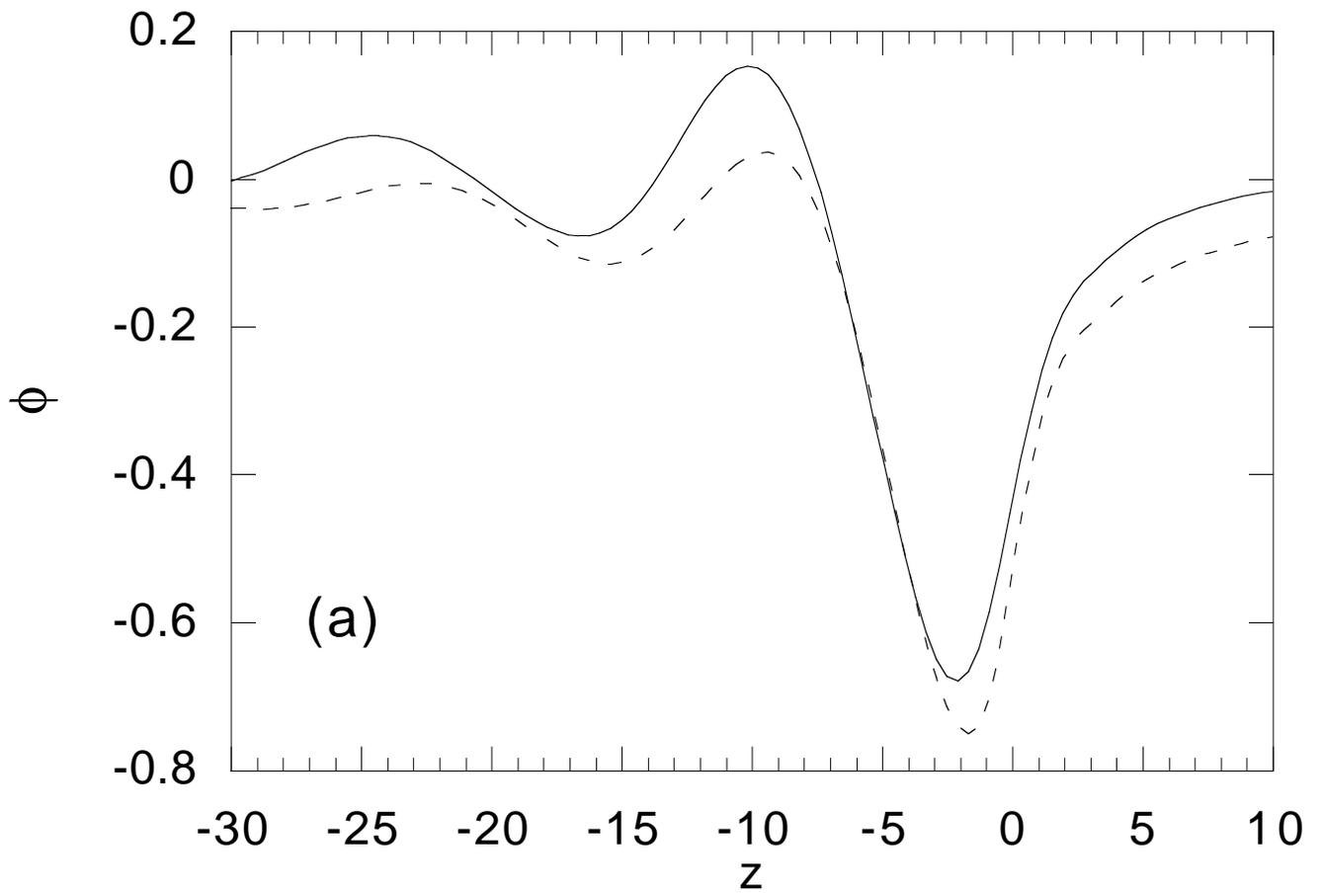
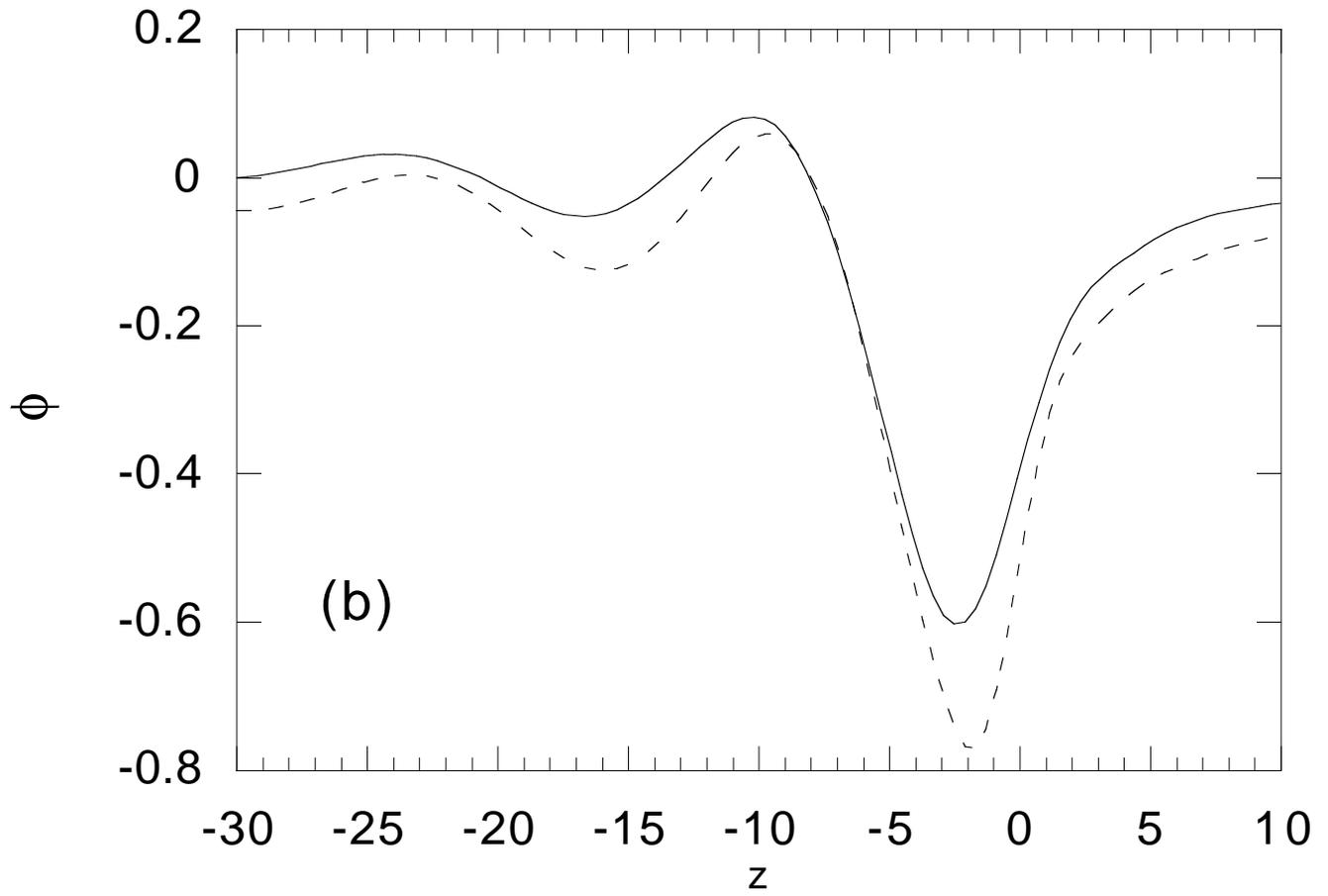
\begin{figure}
\caption[]{
Random potential induced by a proton ($Z_1=1$) moving with $v=1.5$ a.u.
through Al (a), and with $v=1.6$ a.u. through Si (b). 
Solid and dashed lines represent full band-structure and FEG calculations,
respectively.}
\end{figure}

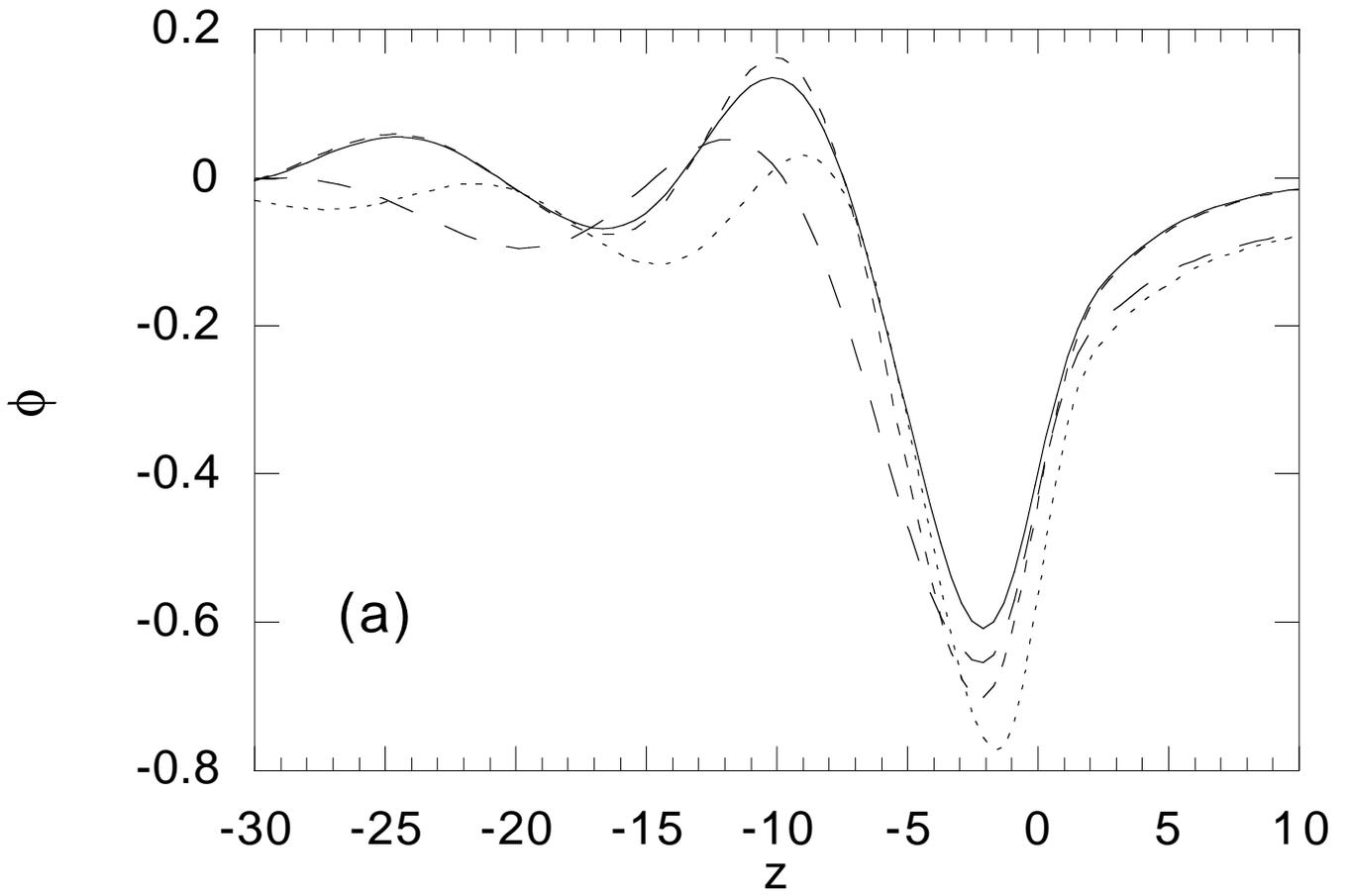
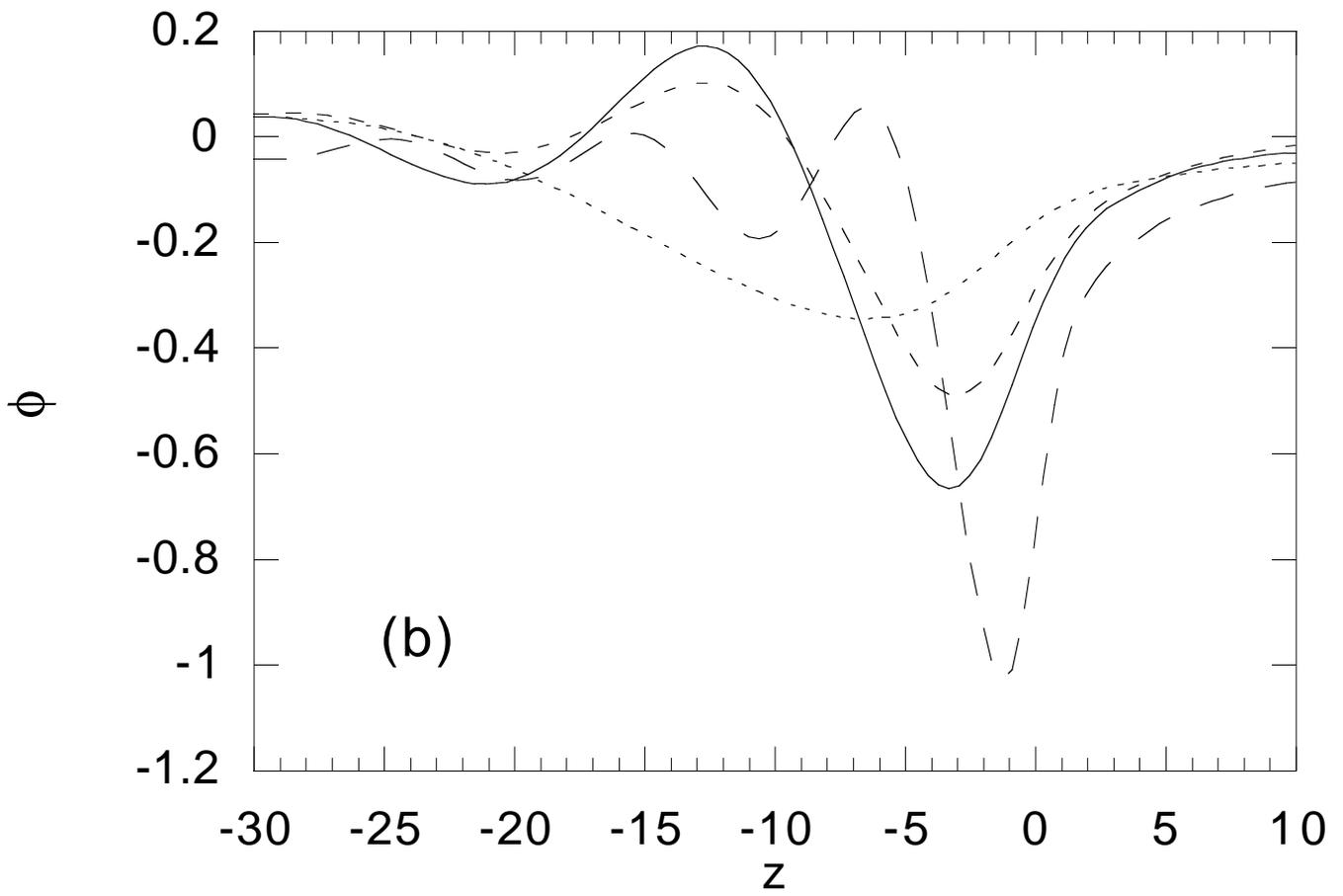
\begin{figure}
\caption[]{
Potential induced by a proton ($Z_1=1$) moving with $v=1.5$ a.u. along the
$\langle 100\rangle$ direction in Al at the impact vector ${\bf
b}=b(0,1,0)$ (a), and with $v=2.0$ a.u. along the $\langle 110\rangle$
direction in Si at the impact vector ${\bf b}=b(-1,1,0)$ (b). Solid and
short-dashed lines represent full band-structure calculations for $b=0$ and
$b=0.25$, respectively. Dashed and dotted lines represent LDA 
calculations for $b=0$ and $b=0.25$, respectively.}
\end{figure}


\begin{references}
\bibitem{note1} The average electron distance is written in terms of the
electron density, as $(1/n_0)=(4\pi/3)r_s^3$.
\bibitem{note2} The use of this so-called LDA should not be confused
with the use, within DFT, of the LDA for the evaluation of
$V_{xc}$ in the Kohn-Sham equation.
\end{references}
\end{document}